\documentclass{article}

\usepackage{arxiv}

\usepackage[utf8]{inputenc} % allow utf-8 input
\usepackage[T1]{fontenc}    % use 8-bit T1 fonts
\usepackage{hyperref}       % hyperlinks
\usepackage{url}            % simple URL typesetting
\usepackage{booktabs}       % professional-quality tables
\usepackage{amsfonts}       % blackboard math symbols
\usepackage{nicefrac}       % compact symbols for 1/2, etc.
\usepackage{microtype}      % microtypography
\usepackage{lipsum}		% Can be removed after putting your text content
\usepackage{graphicx}
\usepackage[numbers,sort&compress]{natbib}
\usepackage{doi}

\title{A Low Complexity and High Modularity Design for Continuously Variable Bandwidth Digital FIR Filter}

\author{ \href{}{\hspace{1mm}Sushmitha Sajeevu} \\
	Department of Electronics and\\
 Communication Engineering\\
	National Institute of Technology Calicut\\
	Kerala, India \\
	Tel.: +91-9645434483\\
 \texttt{ sushmitha\textunderscore p190077ec@nitc.ac.in}\\
	\And
	\href{}{\hspace{1mm}V. Sakthivel} \\
 	Department of Electronics and\\ Communication Engineering\\
 	National Institute of Technology Calicut\\
	Kerala, India \\
	Tel.: +91-9995335962\\
 	\texttt{sakthi517@nitc.ac.in} \\
}

\date{}

\renewcommand{\headeright}{}
\renewcommand{\undertitle}{}
\renewcommand{\shorttitle}{}

 \hypersetup{
 pdftitle={A LOW COMPLEXITY AND HIGH MODULARITY DESIGN FOR CONTINUOUSLY VARIABLE BANDWIDTH DIGITAL FIR FILTER},
 pdfsubject={Signal processing},
 pdfauthor={Sushmitha Sajeevu, V. Sakthivel},
 pdfkeywords={Frequency Response Masking, Pascal structure, Sampling rate converter, Variable bandwidth filter, Software defined radio channelizer},
 }

\begin{document}
\maketitle
\begin{abstract}
	Digital filters with variable bandwidth can be used for a variety of applications.  Arbitrary change in the bandwidth of a digital Finite Impulse Response (FIR) filter can be acquired using sampling rate converters. In this paper, a sampling rate converter is proposed which is generated from Pascal structure, a fractional delay filter having low hardware complexity and high modularity. The proposed sampling rate converter requires lesser number of multipliers for implementation when compared with the sampling rate converters in the literature. A low pass filter having a single bandwidth sandwiched between two sampling rate converters can contribute multiple bandwidths in such a way that each bandwidth is an arbitrary variation of the original bandwidth.  A two stage Frequency  response  masking approach (FRM) is  used for the hardware efficient design of the original low pass filter. A low complexity and high modular novel design for a continuously varying bandwidth of a digital FIR filter is proposed in this paper using the proposed sampling rate converter. The modularity of the Pascal structure  can  be  used  to  control  both  pass  band  ripple  as  well  as  stop  band  attenuation  of the  continuously  variable  bandwidth  FIR  filter  design. Different communication standards in a Software defined radio (SDR) channelizer  is realized using the proposed design of continuously variable bandwidth filter. 
\end{abstract}

% keywords can be removed

\section{Introduction}
Digital filtering is one of the most dynamic tools of digital signal processing. Compared to analog filters, digital filters are less sensitive towards environmental changes, more flexible, programmable and can be easily standardized since they are simply software modules. With the advent of VLSI technology, digital filters are employed in a wide variety of signal processing platforms. Variable bandwidth digital FIR filters have several critical applications in the field of speech signal processing, digital communications, biomedical signal processing etc. \cite{laakso1996splitting,stoyanov1997variable}. Digital channelizer is a prominent part of the digital front end (DFE) in a software defined radio (SDR) based Internet of things (IOT) platform \cite{zeineddine2019design}. The ability to support multiple bands of signal frequencies according to the user demand is a special feature of software defined radio \cite{venosa2011software}. Sharp filtering is necessary to reduce interference from adjacent channels in a digital channelizer. Variable bandwidth digital FIR filter can be used as an efficient digital channelizer.\\ A multitude of approaches has been put forward for the design of variable bandwidth filters \cite{dhabu2020variable}. Variable bandwidth filters can be obtained by varying the filter coefficients. Such variable coefficient filters can be designed by calculating the desired filter coefficient sets in advance and by storing it in the memory \cite{lee1996new} or by calculating the desired filter coefficients on-the-fly \cite{jarske1988simple}. But variable coefficient filter requires huge memory requirement. Hence it cannot be used when the order of the filter is very high.  In fixed coefficient filter design \cite{yu2008low}, without changing the coefficient set, variable bandwidth can be obtained.  Another classification of variable bandwidth filters based on the behaviour of bandwidth variation is, discretely variable bandwidth filters and continuously variable band-width filters. In \cite{mahesh2011low}, \cite{mahesh2008coefficient}, an approach named as Coefficient decimation method(CDM) was developed to vary bandwidth by selective usage of filter coefficients by decimation operation and replacing  other  coefficients  by  zeroes  or  by  discarding  it.  A modified coefficient decimation method(MCDM) was later developed in \cite{ambede2012modified}, in which after decimation, the sign of every alternate retained coefficient is reversed. A combination of CDM and MCDM  was later developed and is termed as improved coefficient decimation method (ICDM) \cite{ambede2015design}. A drawback of the coefficient decimation based methods is that the passband ripple and stopband attenuation in the resultant frequency response characteristics deteriorates as the value of the decimation factor increases. Interpolation as well as Frequency Response Masking (FRM) \cite{sudharman2018design}, \cite{mahesh2008reconfigurable} techniques developed for obtaining sharp transition width filters without increasing the hardware complexity can also be extended to get a variable bandwidth filter. FRM can be extended to a multistage design as in \cite{yu2011mixed} which is called as fast filter bank. However the group delay associated with the fast filter bank based methods is very high. The problem with the FRM based variable bandwidth filters is that it is very difficult to choose the value of the interpolation factor when the desired frequencies are closely spaced. In order to get a better control over the cut-off frequency a combination of the coefficient decimation and interpolation techniques can be used \cite{smitha2009new}. But only a limited number of cut-off frequencies can be obtained using this technique. The all-pass transformation (APT) based variable bandwidth filter is proposed in \cite{mitra1974digital}. Here all the unit delay element in the prototype filter is replaced by an all-pass filter structure of an appropriate order. But the problem with this method is that the variable bandwidth filter is a non-linear-phase filter even if the prototype filter is a linear phase filter.\\
In frequency transformation based methods \cite{oppenheim1976variable}, a variable impulse response filter is obtained by applying frequency transformation on the Taylor expansion of the
impulse response of the prototype filter. Cutoff frequency range of the variable bandwidth filter obtained through this method is very narrow. To improve the cut-off frequency range, frequency transformation techniques  can be combined with coefficient decimation techniques \cite{darak2012design}. Variable bandwidth filters can be obtained through Spectral Parameter Approximation (SPA) techniques \cite{dhabu2017new}. The SPA technique was initially used for designing variable fractional delay filters but later adapted to design variable bandwidth filters. The bandwidth of the filter is varied in discrete steps through these approaches.\\
Continuously varying bandwidth filters refers to the arbitrary variation of the bandwidth. The original bandwidth is altered by a rational value so as to get the desired bandwidth. Continuous variation of bandwidth of a filter can be obtained using an architecture proposed by Harris \cite{harris2009fixed}.  The original single bandwidth low pass filter placed in between two arbitrary interpolators can be used to get multiple bandwidths such that each bandwidth is obtained by arbitrarily changing the original bandwidth. Both bandwidth increase as well as decrease can be done with respect to the fixed filter using this structure \cite{george2014reconfigurable}. Polyphase interpolators were used initially but this has high implementation complexity. Later Polyphase interpolators were replaced by farrow structure \cite{haridas2017low}. But in Farrow structure \cite{farrow1988continuously}, \cite{hermanowicz2006designing}, the computational complexity is proportional to the square of the order of interpolation. In the Farrow structure based continuously variable bandwidth filters, the order of the structure is confined to an even value since the first subfilter is meant to provide a delay. Moreover finding the coefficients of the Farrow subfilter for different orders is a tedious task.\\
 This  proposes a novel design technique for a low hardware complexity, high modularity continuously variable bandwidth digital FIR filter. Major contributions of this paper can be outlined as follows.
 \begin{enumerate}
 \itemindent=-1pt 
     \item A sampling rate converter (SRC) is proposed using Pascal structure, a fractional delay structure.
     \item Using this sampling rate converter and a fixed filter, a continuously variable bandwidth filter is introduced. The fixed filter is implemented using a two stage Frequency response masking technique. This aids to acquire a sharp transition width with reduced complexity.
     \item Pascal structure has got high modularity which is exploited to control passband ripple and stopband attenuation of the variable bandwidth filter.
     \item Hardware complexity in terms of the number of multipliers of the proposed design is compared with the continuously variable bandwidth filters in the literature.
     \item  The channelizer module in a Software defined radio (SDR) is realized using the proposed design of continuously variable bandwidth filters. 
 \end{enumerate}
  
The rest of the paper is organized as follows: Section 2 provides a brief description of the Pascal structure. Section 3 gives the basics of Frequency Response Masking techniques. In Section 4, the proposed
sampling rate converter using Pascal structure is discussed. Section 5 discusses the proposed continuously variable bandwidth filter using the proposed sampling rate converter and two stage FRM. Section 6 includes the design examples and result analysis. Section 7 concludes the paper.
\section{Pascal Structure}
\label{Section 2}
A new structure was introduced in \cite{soontornwong2017low} for implementing a variable delay filter based on Pascal polynomial interpolation. This fractional delay filter can be represented by the equation
\vspace{0.05in}
\begin{equation}
H(z,f)= \sum_{k=0}^{N}P(f,k)(1-z^{-1})^{k} 
\end{equation}
\vspace{0.05in}

where, \(H(z,f)\) is the fractional delay filter, \(f\) is the fractional delay parameter,  \(P(f,k)\) is the Pascal polynomial of k-th degree.
\(P(f,k)\) is defined as:
\vspace{0.05in}
\begin{equation}
P(f,k)=(-1)^{k}\frac{f(f-1)(f-2)...(f-k+1)}{k!}
\end{equation}

\begin{flushleft}
 \(P(f,k)\) can be rewritten as \cite{soontornwong2017low}  
\end{flushleft}
\begin{equation}
   P(f,k)=(1-\frac{f+1}{1})(1-\frac{f+1}{2})(1-\frac{f+1}{3})...(1-\frac{f+1}{k})
  \end{equation}
 \vspace{0.05in} 
  
  Hence the fractional delay filter \(H(z,f)\) can be implemented as in Fig. 1. This structure has got very low complexity than other filter structures in the literature \cite{farrow1988continuously}, \cite{candan2007efficient}. In this structure, same modules are repeated. Hence the structure is highly modular and can be used for modular hardware implementations.
  \begin{figure}[h]
 \centerline{\includegraphics[width=\columnwidth]{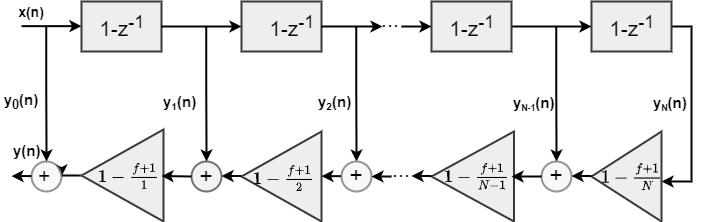}}
 \caption{Pascal Structure for the fractional delay filter \cite{soontornwong2017low}}
 \label{Fig. 1}
 \end{figure}
\section{Frequency Response Masking (FRM) Filter}
\label{Section 3}
The Frequency Response Masking (FRM) approach is a potent method to design a linear phase FIR filter of sharp transition width with reduced complexity. It encapsulates the design of sub-filters, which are cascaded to get an overall sharp transition width filter. It involves interpolation and masking. The interpolation factor L corresponds to the reduction of transition width by L times. The transfer function of a Frequency Response Masking (FRM) filter is given by \cite{lim1986frequency},
\begin{equation}
F(z)=F_{a}(z^{L})F_{ma}(z) +F_{c}(z^{L})F_{mc}(z)
\end{equation}
\begin{equation}
F_{c}(z)=z^{-\frac{(N-1)}{2}} - F_{a}(z)
\end{equation}
where, \(F_{a}(z)\) is the prototype filter or modal filter and \(F_{c}(z)\) is the complementary
filter, L is the interpolation factor and \(F_{ma}(z)\) and \(F_{mc}(z)\) are
the prototype masking filter and complementary masking filter respectively. Fig. 2 shows the basic FRM structure. Fig. 3 shows the illustration of the FRM technique.
 \begin{figure}[h]
 \centerline{\includegraphics[width=8cm]{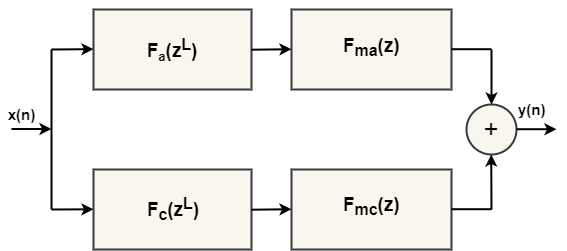}}
 \caption{FRM Structure}
 \label{Fig. 2}
 \end{figure}
  \begin{figure}[h]
 \centerline{\includegraphics[width= 12cm]{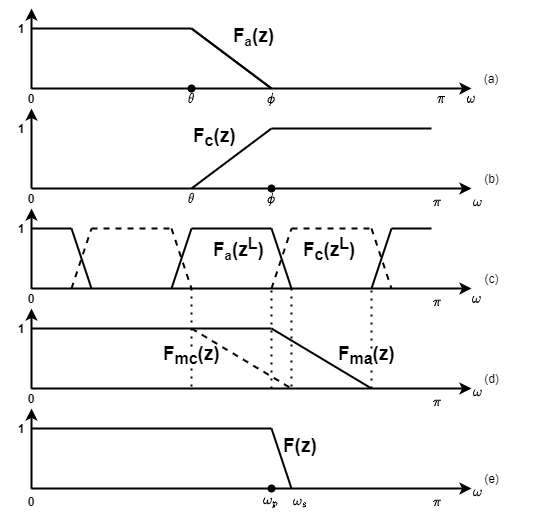}}
 \caption{FRM illustration}
 \label{Fig. 3}
 \end{figure}
 \begin{figure}[h]
 \centerline{\includegraphics[width=\columnwidth]{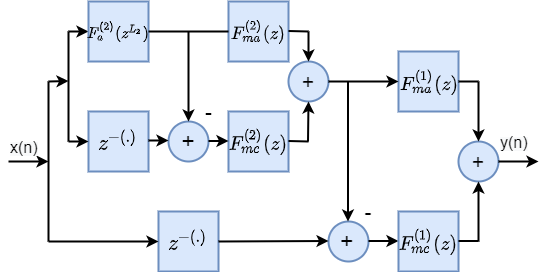}}
 \caption{Two stage FRM structure}
 \label{Fig. 4}
 \end{figure}
\subsection{Two Stage FRM Technique}
Two stage FRM technique is used in cases where we require narrow bandwidth with a hardware computational complexity even lesser than the hardware computational complexity in one stage FRM. Here the modal filter in the FRM is designed using FRM approach. The modal filter can be represented as \cite{lee2006unified},
\begin{equation} 
F^{(1)}_{a}(z)=F^{(2)}_{a}(z^{L_{2}})F^{(2)}_{ma}(z) +F^{(2)}_{c}(z^{L_{2}})F^{(2)}_{mc}(z)
\end{equation} 
where, \(F^{(1)}_{a}(z)\) is the first level modal filter, \(F^{(2)}_{a}(z)\) is the second level modal filter, \(F^{(2)}_{c}(z)\) is the second level complementary filter, \(F^{(2)}_{ma}(z)\) and \(F^{(2)}_{mc}(z)\) are the second level masking filters, \(L_{2}\) is the second level interpolation factor. Fig. 4 shows the two stage FRM structure.
\section{Proposed design of Sampling rate converter using Pascal structure} \label{Section 4}

 \begin{figure}[h]
 \centerline{\includegraphics[width=\columnwidth]{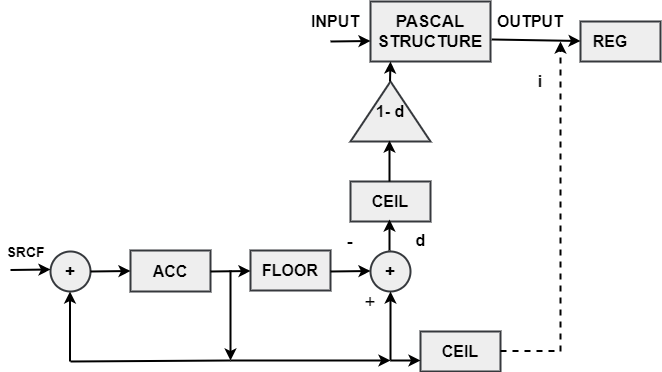}}
 \caption{Proposed Sampling rate converter (SRC)}
 \label{Fig. 5}
 \end{figure}
 
Sampling rate conversion \cite{vaidyanathan2006multirate} is the method of changing the original discrete time samples to another set of time samples for the same continuous time signal. Fractional delay structures are used for delaying the input signal by an arbitrary amount of time interval. Hence the samples at non-integer multiples of the sampling interval can be determined. Pascal structure is a fractional delay structure that has high modularity and low complexity. Fig. 5 shows the proposed design of a sampling rate converter using Pascal structure. The following steps explain the working of the proposed Sampling Rate Converter.
\begin{enumerate}
    \item Initially the accumulator (acc) is loaded with zero.
    \item The output pointer points to the first location (n=0) of the output register.
    \item The fractional part of the accumulator (d) is used to control the fractional delay of the Pascal structure. Explanation is given in the subsection.
    \begin{equation}
    d= acc-\lfloor{acc}\rfloor
    \end{equation}
    \item The ceil of the value in the accumulator,  gives the index(i) of the output of the Pascal structure.
    \begin{equation}
    i=\lceil{acc}\rceil
    \end{equation}
    \item The output at the \(i^{th}\) index of the Pascal structure is saved to the pointed location of the output register.
    \item Output pointer points to the next location of the output register.
    \item Accumulator is accumulated with sampling rate conversion factor( SRCF) value.
    \begin{equation}
    acc = acc + SRCF
    \end{equation}
    \item Repeat steps 3-7 until the value in the accumulator is not greater than the length of the input.
\end{enumerate}
Remarks: Order N=1 is sufficient for proper sampling rate conversion. When the input is having an abrupt change, the order of the Pascal structure can be increased. 
\subsection{Illustration of the Proposed Sampling rate converter}
Following figures illustrates the working of the proposed sampling rate converter. Fig. 6 shows the input signal to a Pascal structure (black colour).  
\begin{figure}[h]
 \centerline{\includegraphics[width=11.5 cm]{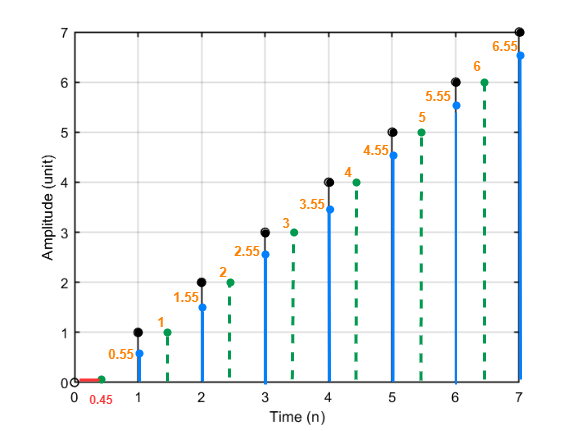}}
 \caption{Input signal and the samples corresponding to a fractional delay of 0.45}
 \label{Fig. 6}
 \end{figure}
 \begin{figure}[h]
 \centerline{\includegraphics[width= 11.5 cm]{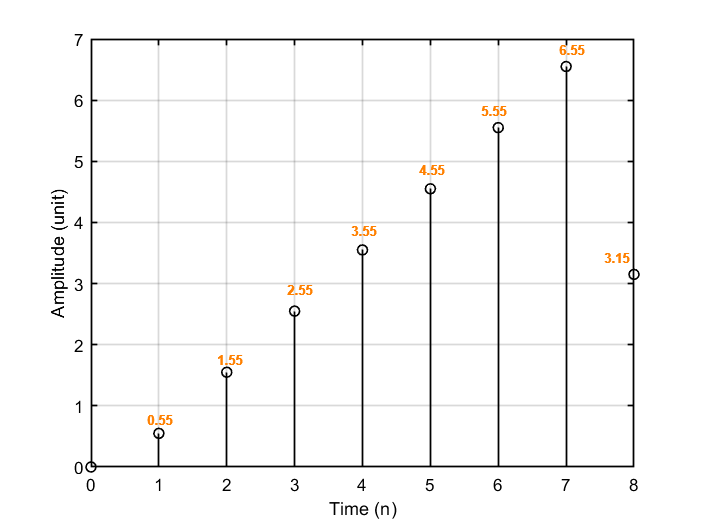}}
 \caption{Output of the Pascal structure when delay is 0.45}
 \label{Fig. 7}
 \end{figure}
 \begin{figure}[h]
 \centerline{\includegraphics[width= 11.5 cm]{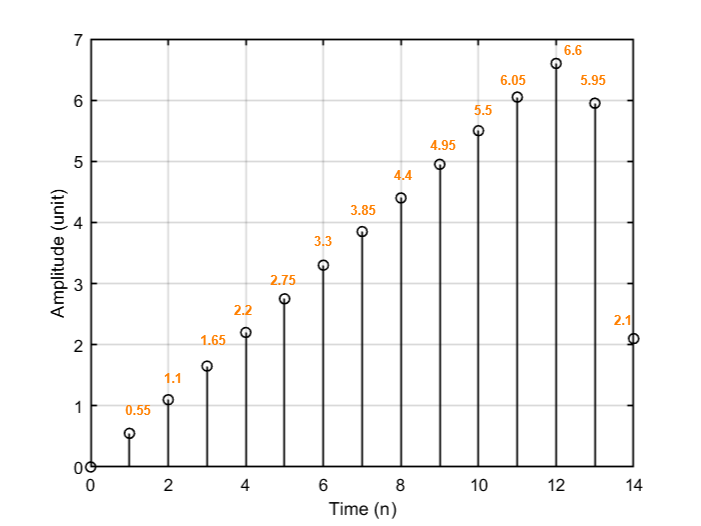}}
 \caption{Output of the sampling rate converter}
 \label{Fig. 8}
 \end{figure}

 \begin{figure}[h]
\centerline{\includegraphics[width=\columnwidth]{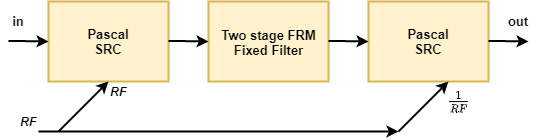}}
\caption{Block diagram of the proposed continuously variable bandwidth FIR filter}
\label{Fig. 9}
 \end{figure}
 If this signal is fractionally delayed by 0.45 unit, then the discrete time samples corresponding to the delayed time cannot be found out readily from the previous samples, since its a fractional delay. We can estimate the values of these samples from the graph. These samples are shown in blue colour. The value of those samples are shown in orange colour. The fractional delay is shown in red colour.\\ The output of the Pascal structure when the fractional delay is 0.45 is shown in Fig. 7. It can be noticed that the fractionally delayed sample values can be obtained through the Pascal structure. Ramp function is used as the input for an easy understanding of the time index. It can be observed from the graph, that if the fractional delay is 0.45, then the samples corresponding to the index \{-0.55,0.55,1.55,2.55...\} will be obtained through the Pascal structure. So to obtain the values corresponding to the index \{-0.45,0.45,1.45,2.45...\}, the delay should be 0.55. The fractional part of the indices of the delayed samples (d) and the fractional delay (f) is related by 
\begin{equation}
f=1-d  \iff d\neq0
\end{equation}
Corresponding to zero delay, the fractional part will be also zero. This is the reason why a ceil block is added before multipier (1-d) in the design of sampling rate converter.   \\
Pascal structure can be used to find the values corresponding to any point in between the samples. That is the main idea which is exploited to design a sampling rate converter. If the sampling rate conversion factor is 2, it implies the values corresponding to the index \{0,2,4,6,...\} is found out. If the sampling rate conversion factor is 0.5, it implies the values corresponding to the index \{0,0.5,1,0.5,1.5,...\} is found out. If the sampling rate conversion factor is 0.55, it implies the values corresponding to the index \{0,0.55,1.1,1.05,2.2,...\} is found out. These indices form an arithmetic progression with the common difference as the sampling rate conversion factor. The sampling rate conversion factor is added over and over again to the accumulator to find out the index values corresponding to a particular sampling rate conversion. The fractional part of the value in the accumulator is separated. Using this value, delay is calculated and the output of the Pascal structure is found. The value corresponding to a particular index is found out by pointing to a particular location in the output of the pascal structure. If we need the value corresponding to the index 0.55, then the ceil of the value in the accumulator (0.55) is calculated. This is the location, the pointer should be pointed. If the fractional part of the value in the accumulator is 0.55, then the delay is 0.45 and the value corresponding to the index 0.55 is at the location 1. This can be observed in Fig. 7. In such a way value corresponding to every index required for sampling rate conversion can be found out. The output of the sampling rate converter is shown in Fig. 8.\\
Remark: Values in the accumulator are the index values corresponding to a particular sampling rate conversion. Hence the fractional part of the value in the accumulator and the fractional part of the indices of the delayed samples (output of the Pascal structure) are same. 
 \section{Proposed design of the continuously variable bandwidth FIR filter}
\label{Section 5}
An efficient method to design a continuously variable bandwidth FIR by keeping the values of the filter coefficients fixed was proposed by Harris \cite{harris2009fixed}. A linear phase digital FIR filter has got the two properties \cite{harris2009fixed}:

\begin{enumerate}
    \item The time interval between the main-lobe peak and first
zero crossing of the impulse response of the filter is the reciprocal of the
filter bandwidth (\(f_{b}\)).
    \item The time interval between samples
is the reciprocal of the sampling rate (\(f_{s}\)).
\end{enumerate}

Hence the number of samples between the main-lobe peak and first zero crossing (\(N_{pz}\)) can be represented as the ratio of sampling rate to filter bandwidth \cite{haridas2017low}.\\
\vspace{0.2 cm}
\begin{equation}
N_{pz}=\frac{f_{s}}{f_{b}}
\end{equation}
\vspace{0.2 cm}

According to this equation, filter bandwidth can be altered by varying the interval between samples by changing the sampling rate. Hence without varying the number of samples between main lobe peak and first zero crossing (\(N_{pz}\)), by changing the sampling rate, filter bandwidth can be varied. An alternate method to alter the filter bandwidth is by  fixing the sampling rate and by altering the number of samples between the main lobe peak and first zero crossing (\(N_{pz}\)). The second approach is used in this proposed design to get a continuously variable bandwidth FIR filter.\\
A fixed filter is there having the original bandwidth. The sampling rate of the input discrete-time signal is changed using the first sampling rate converter and is filtered using the fixed filter. This change \(N_{pz}\) of the original sinc function. The sampling rate of the sinc function is reverted to the initial sampling rate using the second sampling rate converter. So in the overall process, the sampling rate is fixed, whereas \(N_{pz}\) value is changed, which changes the bandwidth accordingly. The block diagram of the proposed continuously variable bandwidth FIR filter is shown in Fig. 9\\
The proposed design of the sampling rate converter discussed in Section 4 can be used for altering the sampling rate. The fixed filter can be implemented using the two stage FRM approach discussed in Section 3. The sampling rate converter is controlled by the sampling rate conversion factor (SRCF). The Reduction factor (RF) is defined  as a ratio of original bandwidth to the desired bandwidth. Original bandwidth is the bandwidth of the original fixed filter. For the first sampling rate converter, the sampling rate conversion factor is the same as the reduction factor and for the second sampling rate converter, the sampling rate conversion factor is the reciprocal of the reduction factor (RF). Unit impulse is given as the input to the first sampling rate converter. After sample rate conversion using the first sampling rate converter, we get an intermediate signal. This intermediate signal is filtered using the fixed two stage FRM filter. The output of the fixed filter, which is a sinc function is reverted to the initial sampling rate using the second sampling rate converter. The frequency response of the final sinc function gives a low pass filter, the bandwidth of which can be controlled by the value of the reduction factor.\\
\begin{figure}[t]
 \centerline{\includegraphics[width=\columnwidth]{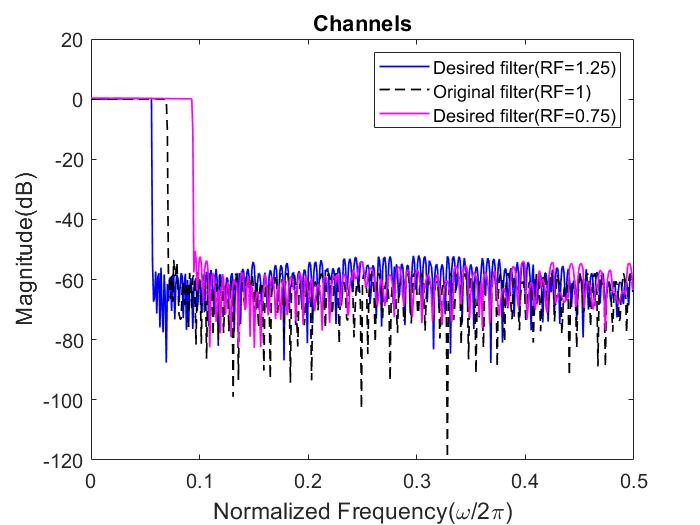}}
 \caption{Variation of bandwidth with reduction factor}
 \label{Fig. 10}
 \end{figure}
The order of the Pascal structure in the sampling rate converter can be updated easily since the Pascal structure design has got high modularity. The proposed design can support a wide dynamic range of bandwidths. The order of the Pascal structure in the two sampling rate converters can be adjusted according to the bandwidth requirements. For applications requiring wide bandwidth, the order of the second sampling rate converter can be kept minimum. Similarly for applications requiring narrow bandwidth, the order of the first sampling rate converter can be kept minimum. Reduced order results in reduced number of multiplications and increased power efficiency. Passband ripples as well as stopband attenuation can also be controlled by changing the order of the Pascal structure in the first sampling rate converter and second sampling rate converter respectively.
 
\section{Results and discussion}
\label{Section 6}
Two examples are discussed in this section to show the modularity and  complexity reduction of the proposed continuously variable bandwidth filter and the application of the proposed filter in a Software defined radio. In  Example-I, the variation of the bandwidth with the change in reduction factor, the variation of the passband ripple and stopband attenuation with the  change  in  the  order  of  the  Pascal  structure in  the sampling rate converter for a specific bandwidth, the variation of the order of the pascal structure in the sampling rate converter for different bandwidths for a specific passband ripple and stop band attenuation, hardware complexity calculation in terms of the number of multipliers etc. are discussed. In  Example-II,  the  proposed  scheme  is  used  for designing the channelizer module in the Software  defined  radio.   A software defined radio for 11 communication standards is designed using the proposed design and  is  compared  in  terms  of  hardware  complexity  with  the  existing  works  in  the  literature.
 \begin{figure}[t]
 \centerline{\includegraphics[width=\columnwidth]{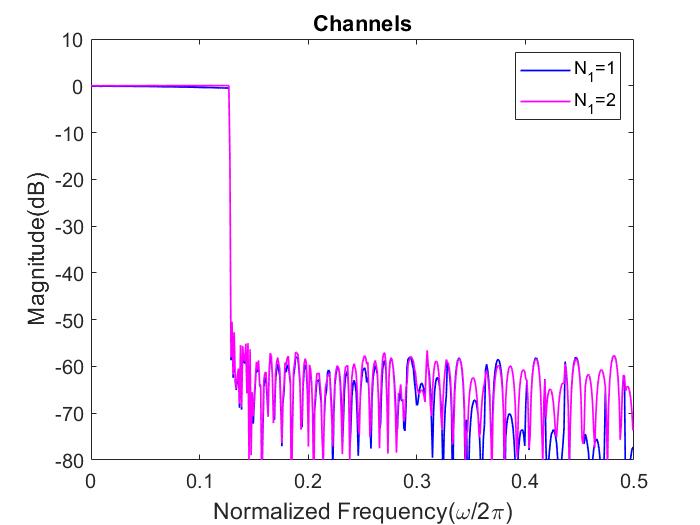}}
 \caption{Variation of Passband ripple with \(N_{1}\)}
 \label{Fig. 11}
 \end{figure}
\begin{figure}[t]
 \centerline{\includegraphics[width=\columnwidth]{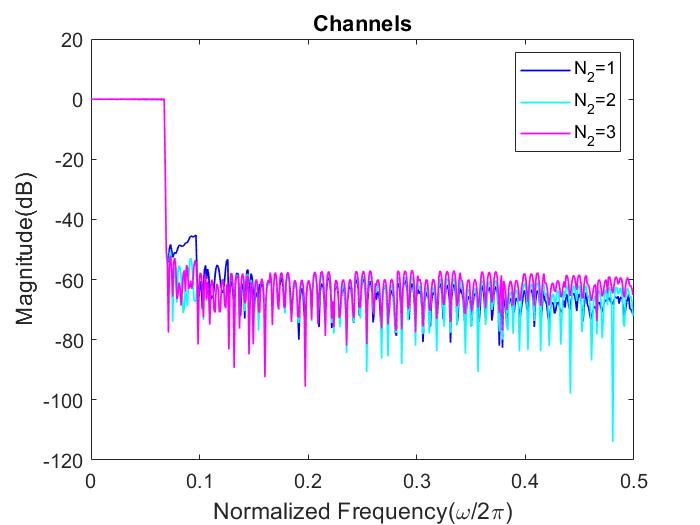}}
 \caption{Variation of Stop band attenuation with \(N_{2}\)}
 \label{Fig. 12}
 \end{figure}
 \begin{figure}[t]
 \centerline{\includegraphics[width=\columnwidth]{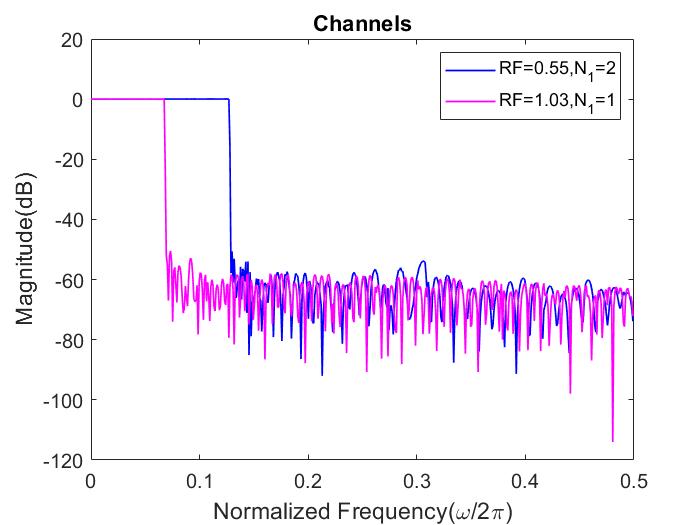}}
 \caption{\(N_{1}\) variations with RF for same passband ripple}
 \label{Fig. 13}
 \end{figure}
 \begin{figure}[t]
 \centerline{\includegraphics[width=\columnwidth]{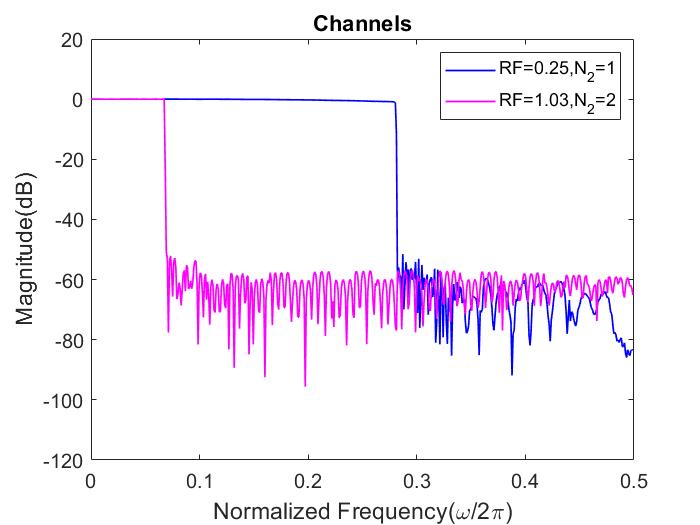}}
 \caption{\(N_{2}\) variations with RF for nearly same stopband attenuation}
 \label{Fig. 14}
 \end{figure}

\subsection{Example I}

The fixed filter in the proposed design is designed for the following specification,\\
\emph{Passband edge frequency}: \(0.14\pi\)\\
\emph{Stopband edge frequency}: \(0.141\pi\)\\
\emph{Maximum passband ripple}:     0.0298 dB\\
\emph{Minimum stopband attenuation}: 50 dB\\

 The fixed filter will give a single bandwidth which is marked as the bandwidth 
 of the original filter in Fig. 10. Using the proposed design we can get an increased bandwidth as well as decreased bandwidth with respect to the original bandwidth. For a reduction factor greater than unity, a decreased bandwidth can be obtained. Similarly, for a reduction factor lesser than unity, an increased bandwidth can be obtained. The fixed filter is accompanied by two sampling rate converters at either side in the proposed design. By changing the reduction factor values, an increase or decrease in the bandwidth can be obtained. This is shown in Fig. 10. The sampling rate converter has a Pascal structure. One of the special features of the Pascal structure is its modularity. This feature can be exploited to control passband ripples as well as stop band attenuation. It has been observed that the order of the Pascal structure in the first sampling rate converter \(N_{1}\) has a direct control over the passband ripples. Similarly, the order of the Pascal structure in the second sampling rate converter \(N_{2}\) has a direct control over the stopband attenuation. This is shown in Fig. 11 and Fig. 12 respectively. In Fig. 11, for a reduction factor of 0.55, when \(N_{1}\) is 1, the passband ripple is 0.04 dB. This can be improved by  increasing the \(N_{1}\) to 2. When  \(N_{1}\) is changed to 2, the passband  ripple is 0.02 dB. In Fig. 12, for a reduction factor of 1.03, when \(N_{2}\) is 1, the stopband attenuation is 45.49 dB. When \(N_{2}\) is increased to 2, the stopband attenuation is improved to 50.47 dB. When \(N_{2}\) is increased to 3, the stopband attenuation is improved to 52.12 dB. Thereafter increasing \(N_{2}\), does not results in improvement in stopband attenuation. In such a way, minimum stopband attenuation can be controlled by altering the value of \(N_{2}\).\\

 For a narrow bandwidth frequency response (RF high), the \(N_{1}\) required to attain a particular passband ripple is comparatively less when compared to a wide bandwidth frequency response (RF low). This is shown in Fig. 13. To attain a maximum passband ripple of 0.02 dB,  \(N_{1}\) should be 2 when RF is 0.55. But for attaining the same for RF is 1.03, the value of \(N_{1}\) need to be only 1. For a wide bandwidth frequency response (RF low), the \(N_{2}\) required to attain a particular stopband attenuation is comparatively less when compared to a narrow bandwidth frequency response (RF high). This is shown in Fig. 14. To attain a minimum stopband attenuation of 50.47 dB, \(N_{2}\) should be 2 when RF is 1.03. But for attaining a stopband attenuation of 51.56 dB, 1 is enough as \(N_{2}\). The change in the order of the Pascal structure can be easily done due to the modularity of the Pascal structure.\\
 \begin{table}[h]
\centering
\caption{Coefficients in one stage FRM}
\begin{tabular}{ l l l l }
 \hline
 \(F_{a}(z)\) & \(F_{ma}(z)\) & \(F_{mc}(z)\) 
 \\ [1.25ex] 
 \hline
 229 & 152 & 152 \\ [1ex] 
  \hline
\end{tabular}
\label{Table 1}
\end{table}

\begin{table}[h]
\centering
\caption{Multipliers in one stage FRM}
\begin{tabular}{ l l l l l}
 \hline
 \(F_{a}(z)\) & \(F_{ma}(z)\) & \(F_{mc}(z)\) & Total
 \\ [1.25ex] 
 \hline
 115 & 76 & 76 & 267 \\ [1ex] 
  \hline
\end{tabular}
\label{Table 2}
\end{table}

\begin{table}[h!]
\centering
\caption{Coefficients in two stage FRM}
\begin{tabular}{ l l l l l l }
 \hline
 \(F^{(2)}_{a}(z)\) & \(F^{(2)}_{ma}(z)\) & \(F^{(2)}_{mc}(z)\) & \(F^{(1)}_{ma}(z)\) & \(F^{(1)}_{mc}(z)\) 
 \\ [1.25ex] 
 \hline
 87 & 24 & 24 & 152 & 152 \\ [1ex]

 \hline
\end{tabular}
\label{Table 3}
\end{table}

\begin{table}[h!]
\centering
\caption{Multipliers in two stage FRM}
\begin{tabular}{ l l l l l l l }
 \hline
 \(F^{(2)}_{a}(z)\) & \(F^{(2)}_{ma}(z)\) & \(F^{(2)}_{mc}(z)\) & \(F^{(1)}_{ma}(z)\) & \(F^{(1)}_{mc}(z)\) & Total
 \\ [1.25ex] 
 \hline
 44 & 12 & 12 & 76 & 76 & 220 \\ [1ex]

 \hline
\end{tabular}
\label{Table 4}
\end{table}

\begin{table}[h!]
\centering
\caption{Number of multiplications versus order in Pascal structure}
\begin{tabular}{ l l l l }
 \hline
  & \(N_{p}=1\)  & \(N_{p}=2\) & \(N_{p}\geq 3\) 
 \\ [1.25ex] 
 \hline
 Pascal  & 1 & 3 &  \(2N_{p}-1\)\\ [1.5ex]
\hline
\end{tabular}
\label{Table 5}
\end{table}

 \begin{table}[h!]
\centering
\caption{Hardware complexity comparison of sampling rate converters}
\begin{tabular}{ l l l l l l l}
 \hline
  & \(N_{s}=1\)  & \(N_{s}=2\) & \(N_{s}=3\) & \(N_{s}=4\) & \(N_{s}=5\) & \(N_{s}=6\)
 \\ [1.25ex] 
 \hline
 Farrow  & - & 5 & - & 15 & - & 53\\ [1.5ex]
 Pascal  & 2 & 4 & 6 & 8 & 10 & 12\\ [1.5ex]
\hline
\end{tabular}
\label{Table 6}
\end{table}

 In the proposed design, a two stage FRM is used for the design of the fixed filter. The computational complexity of the two stage FRM is lesser than that of one stage FRM. Table 1 shows the number of coefficients in one stage FRM for the particular fixed filter specification used in this example. Since the filters in the FRM design are symmetric, the number of multipliers required will be half of the number of coefficients. This is shown in Table 2. Table 3 shows the coefficients in two stage FRM corresponding to every filter in the two stage FRM design for the fixed filter specification. Table 4 shows the multipliers in two stage FRM. It can be observed that the number of multipliers required for two stage FRM is lesser than one stage FRM. 
 Pascal structure is used in the design of sampling rate converter.The number of multiplications versus order in the Pascal structure is shown in Table 5. An additional multiplier is used in the design of the sampling rate converter. Hardware complexity comparison of Pascal structure based sampling rate converter in terms of  the number of multipliers with Farrow structure based sampling rate converter which is used in \cite{haridas2017low} is shown in Table 6. For farrow structure based sampling rate converter, only even order is possible. But in Pascal structure based sampling rate converter every order is possible and moreover order can be changed on the fly due to the modularity of the structure. It can be observed from Table 6 that the number of multipliers required for Pascal structure based sampling rate converter is lesser than that of Farrow based sampling rate converter. The number of multipliers in the SRC for different RF values used in this example is shown in Table 7. Hardware complexity comparison in terms of number of multipliers of the existing continuously varying bandwidth filters with the proposed method for the specification in this example is shown in Table 8. It can be observed that the hardware complexity of the proposed method is lesser than the previous methods in the literature.

 \begin{table}[]
\centering
\caption{Number of multipliers in the SRC for different RF in Example I}
\begin{tabular}{ l l l l }
 \hline
\(RF\) & \(N_{1}\) & \(N_{2}\) & \( SRC_{mult}\)  \\ [1ex] 
 \hline
 0.25 & 2 & 1 & 6\\ 
 0.55 & 2 & 2 & 8 \\
0.75 & 2 & 2 & 8\\
 1.03 & 1 & 3 & 8\\
 1 & 1 & 3 & 8 \\
 1.25 & 1 & 3 & 8\\
  \hline
\end{tabular}
\label{Table 7}
\end{table}
\begin{table}[h!]
\centering
\caption{Hardware complexity comparison in terms of number of multipliers}
\begin{tabular}{ l l l l}
 \hline
 Design Method & Fixed filter & SRC & Total\\ [1ex] 
 \hline
 Harris et al.\cite{harris2009fixed} & 559 & 1920 & 2479\\[1ex]  
 James et al.\cite{george2014reconfigurable} & 267 & 1920 & 2187\\[1ex]  
 Nisha et al.\cite{haridas2017low} & 267 & 20 & 287 \\[1ex] 
 Proposed Method & 220 & 8 & 228\\[1ex] 
 
 \hline
\end{tabular}
\label{Table 8}
\end{table}

\subsection{Example II}
The proposed continuously variable bandwidth filter can be designed for a variety of applications. The channelizer in the Software Defined Radio (SDR) that can support multiple channels of different wireless standards can be efficiently designed using the proposed scheme. A set of 11 wireless communication standards and their respective bandwidths are provided in Table 9. \cite{haridas2017low},\cite{sakthivel2018low}.\\
In an SDR, wireless channels should have minimum interference from neighbouring channels. It is advisable to have sharp transition width. Hence Frequency response masking (FRM) approach is used. The two stage FRM approach explained in Section 3 is used for minimum hardware overhead. 
The proposed design is done using the following fixed filter specification,\\
\emph{Passband edge frequency}: \(0.18\pi\)\\
\emph{Stopband edge frequency}: \(0.181\pi\)\\
\emph{Maximum passband ripple}:     0.02 dB\\
\emph{Minimum stopband attenuation}: 50 dB\\

\begin{table}[h!]
\centering
\caption{Standards of wireless technologies}
\begin{tabular}{ l l l }
 \hline
Wireless & Bandwidth (MHz) & Normalised bandwidth
 \\ Standards & & (10 MHz) \\ [1ex] 
 \hline
 4G MBWA & 0.625 & 0.0625\\ 
 ADSL 2 & 0.962 & 0.0962 \\
 IEEE 802.15.1(Bluetooth) & 1 & 0.1\\
 CDMA 2000 1x & 1.25 & 0.125 \\
 Digital radio & 1.712 & 0.1712 \\
 IEEE 802.16d 1x & 1.75 & 0.175\\
 IEEE 802.15.4a (Zigbee)& 2 & 0.2 \\
 ADSL 2+ & 2.109 & 0.2109 \\
 CDMA 2000 2x & 2.5 & 0.25 \\
 IEEE 802.16d 2x & 3.5 & 0.35 \\
 CDMA 2000 3x & 3.75 & 0.375\\
 \hline
\end{tabular}
\label{Table 9}
\end{table}
A reconfigurable filter is realized using the proposed design that can switch between various communication standards given in Table 9, with this fixed filter at the middle and two sampling rate converters at either side.\\

Table 10 shows the hardware complexity calculation in terms of the number of multipliers for the sampling rate converter \(SRC_{mult}\) for different bandwidths of the channelizer module. The hardware complexity calculation in terms of number of multipliers of the proposed continuously variable bandwidth filter is shown in Table 11.  The resultant frequency response is shown in Fig. 15. The proposed reconfigurable design is compared with the state of the art. The hardware complexity comparison in terms of the number of multipliers is shown in Table 12. The number of multipliers is taken into account since multiplications contribute the maximum computational complexity in DSP applications. Hence it can be observed that the proposed design has the least hardware complexity.

%\begin{table*}[width=1.0\textwidth,cols=4,pos=t]
\begin{table*}
	\caption{Hardware complexity calculation of the sampling rate converters}
	\begin{tabular}{lllllll}
		\toprule
		Normalised Bandwidth & \(RF\) & \( N_{1}\) & \( N_{2}\) & Passband ripple (dB) & Stop band attenuation (dB) & \(SRC_{mult}\) \\
		\midrule
		0.0625 & 1.44 & 1 & 4  & 0.0012 & 55.41 & 10 \\
		0.0962 & 0.9356 & 1 & 3 & 0.005 & 55.5 & 8  \\
		0.1 & 0.9 & 1 & 3 & 0.008 & 51.97 & 8 \\
		0.125 & 0.72 & 2 & 3 & 0.009 & 52.13 & 10 \\
		0.1712 & 0.5257 & 2 & 3 & 0.012 & 52.01 & 10  \\
		0.175 & 0.5143 & 2 & 3 & 0.02 & 53.11 & 10  \\
		0.2 & 0.45 & 2 & 2 & 0.08 & 52.23 & 8  \\
		0.2109 & 0.4267 & 2 & 2 & 0.1 & 50.78 & 8  \\
		0.25 & 0.36 & 2 & 2 & 0.19 & 52.69 & 8  \\
		0.35 & 0.2571 & 3 & 2 & 0.2 & 50.35 & 10  \\
		0.375 & 0.24 & 3 & 2 & 0.2 & 51.64 & 10  \\
		\bottomrule
	\end{tabular}
	\label{Table 10}
\end{table*}
\begin{figure}
\centerline{\includegraphics[width=\columnwidth]{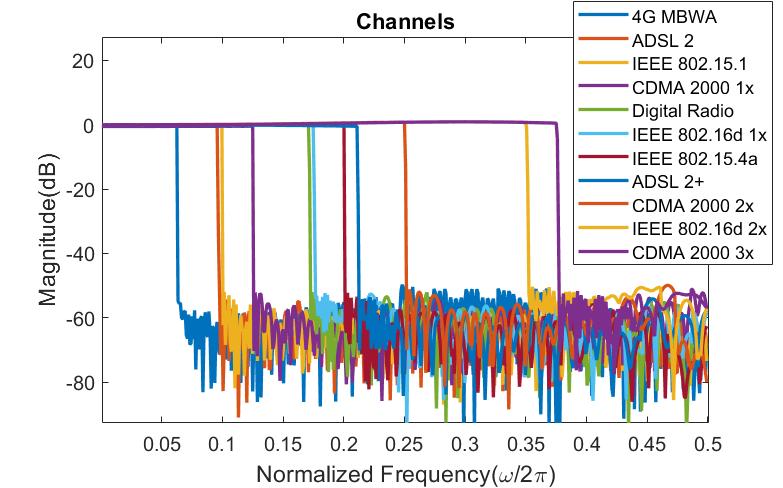}}
\caption{Frequency response of the SDR Channelizer}
\label{Fig. 15}
\end{figure}
  \begin{table}[h!]
\centering
\caption{Hardware complexity Calculation in terms of number of multipliers}
\begin{tabular}{ l l l l l l l l}
 \hline
 \(F^{(2)}_{a}(z)\) & \(F^{(2)}_{ma}(z)\) & \(F^{(2)}_{mc}(z)\) & \(F^{(1)}_{ma}(z)\) & \(F^{(1)}_{mc}(z)\) & SRC & Total
 \\ [1.25ex] 
 \hline
 47 & 14 & 14 & 80 & 80 & 10 & 245\\ [1ex] 
 \hline
\end{tabular}
\label{Table 11}
\end{table}

 \begin{table}[h!]
\centering
\caption{Hardware complexity comparison in terms of number of multipliers}
\begin{tabular}{ l l l l}
 \hline
 Design Method & Fixed filter & SRC & Total\\ [1ex] 
 \hline
 James et al.\cite{george2014reconfigurable} & 281 & 1920 & 2201\\[1ex]  
 Nisha et al.\cite{haridas2017low} & 281 & 30 & 311 \\[1ex] 
 Proposed Method & 235 & 10 & 245\\[1ex] 
 \hline
\end{tabular}
\label{Table 12}
\end{table}

\section{Conclusion} \label{Section 7}

In this paper, a novel low hardware complexity and highly modular design for continuously variable bandwidth digital FIR filter is proposed. The hardware complexity of the proposed design in terms of the number of multipliers is analyzed. The proposed design has got a significant reduction in hardware complexity. The channelizer module in the Software defined radio is realized using the proposed design. Bandwidths corresponding to multiple communication standards are derived from a fixed original filter and two sampling rate converters. The design is highly reconfigurable due to the modularity of the pascal structure.

\bibliographystyle{elsarticle-num}
\bibliography{ref}  %%% Uncomment this line and comment out the ``thebibliography'' section below to use the external .bib file (using bibtex) .

\begin{thebibliography}{10}
\expandafter\ifx\csname url\endcsname\relax
  \def\url#1{\texttt{#1}}\fi
\expandafter\ifx\csname urlprefix\endcsname\relax\def\urlprefix{URL }\fi
\expandafter\ifx\csname href\endcsname\relax
  \def\href#1#2{#2} \def\path#1{#1}\fi

\bibitem{laakso1996splitting}
T.~I. Laakso, V.~Valimaki, M.~Karjalainen, U.~K. Laine,
  \href{https://ieeexplore.ieee.org/document/482137}{Splitting the unit delay
  [fir/all pass filters design]}, IEEE Signal Processing Magazine 13~(1) (1996)
  30--60.
\newline\urlprefix\url{https://ieeexplore.ieee.org/document/482137}

\bibitem{stoyanov1997variable}
G.~Stoyanov, M.~Kawamata,
  \href{https://citeseerx.ist.psu.edu/viewdoc/download?doi=10.1.1.1065.5144&rep=rep1&type=pdf}{Variable
  digital filters}, J. Signal Processing 1~(4) (1997) 275--289.
\newline\urlprefix\url{https://citeseerx.ist.psu.edu/viewdoc/download?doi=10.1.1.1065.5144&rep=rep1&type=pdf}

\bibitem{zeineddine2019design}
A.~Zeineddine,
  \href{https://tel.archives-ouvertes.fr/tel-02877254/document}{Design of a
  generic digital front-end for the internet of things}, Ph.D. thesis,
  CentraleSup{\'e}lec (2019).
\newline\urlprefix\url{https://tel.archives-ouvertes.fr/tel-02877254/document}

\bibitem{venosa2011software}
E.~Venosa, F.~A. Palmieri, et~al., Software radio: sampling rate selection,
  design and synchronization, Springer Science \& Business Media, 2011.

\bibitem{dhabu2020variable}
S.~Dhabu, A.~Ambede, N.~Agrawal, K.~Smitha, S.~Darak, A.~Vinod,
  \href{https://link.springer.com/article/10.1007/s42452-020-2140-6}{Variable
  cutoff frequency fir filters: A survey}, SN Applied Sciences 2~(3) (2020)
  1--23.
\newline\urlprefix\url{https://link.springer.com/article/10.1007/s42452-020-2140-6}

\bibitem{lee1996new}
H.-R. Lee, C.-W. Jen, C.-M. Liu,
  \href{https://ieeexplore.ieee.org/document/536760?reload=trueLee}{A new
  hardware-efficient architecture for programmable fir filters}, IEEE
  Transactions on Circuits and Systems II: Analog and Digital Signal Processing
  43~(9) (1996) 637--644.
\newline\urlprefix\url{https://ieeexplore.ieee.org/document/536760?reload=trueLee}

\bibitem{jarske1988simple}
P.~Jarske, Y.~Neuvo, S.~K. Mitra,
  \href{https://www.sciencedirect.com/science/article/abs/pii/0165168488900904}{A
  simple approach to the design of linear phase fir digital filters with
  variable characteristics}, Signal Processing 14~(4) (1988) 313--326.
\newline\urlprefix\url{https://www.sciencedirect.com/science/article/abs/pii/0165168488900904}

\bibitem{yu2008low}
Y.~J. Yu, Y.~C. Lim, D.~Shi,
  \href{https://ieeexplore.ieee.org/document/4696049}{Low-complexity design of
  variable bandedge linear phase fir filters with sharp transition band}, IEEE
  Transactions on Signal Processing 57~(4) (2008) 1328--1338.
\newline\urlprefix\url{https://ieeexplore.ieee.org/document/4696049}

\bibitem{mahesh2011low}
R.~Mahesh, A.~P. Vinod, \href{https://doi.org/10.1049/iet-cds.2010.0010}{Low
  complexity flexible filter banks for uniform and non-uniform channelisation
  in software radios using coefficient decimation}, IET Circuits, Devices \&
  Systems 5~(3) (2011) 232--242.
\newline\urlprefix\url{https://doi.org/10.1049/iet-cds.2010.0010}

\bibitem{mahesh2008coefficient}
R.~Mahesh, A.~P. Vinod,
  \href{https://ieeexplore.ieee.org/document/4541359}{Coefficient decimation
  approach for realizing reconfigurable finite impulse response filters}, in:
  2008 IEEE international symposium on circuits and systems, IEEE, 2008, pp.
  81--84.
\newline\urlprefix\url{https://ieeexplore.ieee.org/document/4541359}

\bibitem{ambede2012modified}
A.~Ambede, K.~G. Smitha, A.~P. Vinod,
  \href{https://ieeexplore.ieee.org/document/6256379}{A modified coefficient
  decimation method to realize low complexity fir filters with enhanced
  frequency response flexibility and passband resolution}, in: 2012 35th
  international conference on telecommunications and signal processing (TSP),
  IEEE, 2012, pp. 658--661.
\newline\urlprefix\url{https://ieeexplore.ieee.org/document/6256379}

\bibitem{ambede2015design}
A.~Ambede, S.~Shreejith, A.~P. Vinod, S.~A. Fahmy,
  \href{https://ieeexplore.ieee.org/document/7206539}{Design and realization of
  variable digital filters for software-defined radio channelizers using an
  improved coefficient decimation method}, IEEE Transactions on Circuits and
  Systems II: Express Briefs 63~(1) (2015) 59--63.
\newline\urlprefix\url{https://ieeexplore.ieee.org/document/7206539}

\bibitem{sudharman2018design}
S.~Sudharman, T.~Bindiya,
  \href{https://ieeexplore.ieee.org/document/8584138}{Design of power efficient
  variable bandwidth non-maximally decimated frm filters for wideband
  channelizer}, IEEE Transactions on Circuits and Systems II: Express Briefs
  66~(9) (2018) 1597--1601.
\newline\urlprefix\url{https://ieeexplore.ieee.org/document/8584138}

\bibitem{mahesh2008reconfigurable}
R.~Mahesh, A.~P. Vinod,
  \href{https://ieeexplore.ieee.org/document/4469990}{Reconfigurable frequency
  response masking filters for software radio channelization}, IEEE
  Transactions on Circuits and Systems II: Express Briefs 55~(3) (2008)
  274--278.
\newline\urlprefix\url{https://ieeexplore.ieee.org/document/4469990}

\bibitem{yu2011mixed}
Y.~J. Yu, W.~J. Xu,
  \href{https://ieeexplore.ieee.org/document/6031936}{Mixed-radix fast filter
  bank approach for the design of variable digital filters with simultaneously
  tunable bandedge and fractional delay}, IEEE transactions on signal
  processing 60~(1) (2011) 100--111.
\newline\urlprefix\url{https://ieeexplore.ieee.org/document/6031936}

\bibitem{smitha2009new}
K.~G. Smitha, A.~P. Vinod,
  \href{https://www.sciencedirect.com/science/article/abs/pii/S1874490709000214?via3Dihub}{A
  new low power reconfigurable decimation--interpolation and masking based
  filter architecture for channel adaptation in cognitive radio handsets},
  Physical Communication 2~(1-2) (2009) 47--57.
\newline\urlprefix\url{https://www.sciencedirect.com/science/article/abs/pii/S1874490709000214?via3Dihub}

\bibitem{mitra1974digital}
S.~Mitra, K.~Hirano,
  \href{https://ieeexplore.ieee.org/document/1083908}{Digital all-pass
  networks}, IEEE Transactions on Circuits and Systems 21~(5) (1974) 688--700.
\newline\urlprefix\url{https://ieeexplore.ieee.org/document/1083908}

\bibitem{oppenheim1976variable}
A.~Oppenheim, W.~Mecklenbrauker, R.~Mersereau,
  \href{https://ieeexplore.ieee.org/document/1084202}{Variable cutoff linear
  phase digital filters}, IEEE Transactions on circuits and systems 23~(4)
  (1976) 199--203.
\newline\urlprefix\url{https://ieeexplore.ieee.org/document/1084202}

\bibitem{darak2012design}
S.~J. Darak, A.~P. Vinod, E.~M. Lai,
  \href{https://ieeexplore.ieee.org/abstract/document/6271999}{Design of
  variable linear phase fir filters based on second order frequency
  transformations and coefficient decimation}, in: 2012 IEEE International
  Symposium on Circuits and Systems (ISCAS), IEEE, 2012, pp. 3182--3185.
\newline\urlprefix\url{https://ieeexplore.ieee.org/abstract/document/6271999}

\bibitem{dhabu2017new}
S.~Dhabu, A.~P. Vinod,
  \href{https://link.springer.com/article/10.1007/s00034-016-0407-3}{A new
  time-domain approach for the design of variable fir filters using the
  spectral parameter approximation technique}, Circuits, Systems, and Signal
  Processing 36~(5) (2017) 2154--2165.
\newline\urlprefix\url{https://link.springer.com/article/10.1007/s00034-016-0407-3}

\bibitem{harris2009fixed}
F.~Harris, \href{https://ieeexplore.ieee.org/document/5172575}{Fixed length fir
  filters with continuously variable bandwidth}, in: 2009 1st International
  Conference on Wireless Communication, Vehicular Technology, Information
  Theory and Aerospace \& Electronic Systems Technology, IEEE, 2009, pp.
  931--935.
\newline\urlprefix\url{https://ieeexplore.ieee.org/document/5172575}

\bibitem{george2014reconfigurable}
J.~T. George, E.~Elias,
  \href{https://www.sciencedirect.com/science/article/abs/pii/S1434841113002641}{Reconfigurable
  channel filtering and digital down conversion in optimal csd space for
  software defined radio}, AEU-International Journal of Electronics and
  Communications 68~(4) (2014) 312--321.
\newline\urlprefix\url{https://www.sciencedirect.com/science/article/abs/pii/S1434841113002641}

\bibitem{haridas2017low}
N.~Haridas, E.~Elias,
  \href{https://ietresearch.onlinelibrary.wiley.com/doi/10.1049/iet-spr.2016.0055}{Low-complexity
  technique to get arbitrary variation in the bandwidth of a digital fir
  filter}, IET Signal Processing 11~(4) (2017) 372--377.
\newline\urlprefix\url{https://ietresearch.onlinelibrary.wiley.com/doi/10.1049/iet-spr.2016.0055}

\bibitem{farrow1988continuously}
C.~W. Farrow, \href{https://ieeexplore.ieee.org/document/15483}{A continuously
  variable digital delay element}, in: 1988., IEEE International Symposium on
  Circuits and Systems, IEEE, 1988, pp. 2641--2645.
\newline\urlprefix\url{https://ieeexplore.ieee.org/document/15483}

\bibitem{hermanowicz2006designing}
E.~Hermanowicz, \href{https://ieeexplore.ieee.org/document/7071692}{Designing
  linear-phase digital differentiators a novel approach}, in: 2006 14th
  European Signal Processing Conference, IEEE, 2006, pp. 1--4.
\newline\urlprefix\url{https://ieeexplore.ieee.org/document/7071692}

\bibitem{soontornwong2017low}
P.~Soontornwong, T.-B. Deng, S.~Chivapreecha,
  \href{https://ieeexplore.ieee.org/document/8027082/citations?tabFilter=papers#citations}{Low-complexity
  and high-modularity structure for implementing transient-free pascal-delay
  filter}, IEEE Transactions on Signal Processing 65~(23) (2017) 6233--6243.
\newline\urlprefix\url{https://ieeexplore.ieee.org/document/8027082/citations?tabFilter=papers#citations}

\bibitem{candan2007efficient}
C.~Candan, \href{https://ieeexplore.ieee.org/abstract/document/4035693}{An
  efficient filtering structure for lagrange interpolation}, IEEE Signal
  Process. Lett. 14~(1) (2007) 17--19.
\newline\urlprefix\url{https://ieeexplore.ieee.org/abstract/document/4035693}

\bibitem{lim1986frequency}
Y.~Lim, \href{https://ieeexplore.ieee.org/document/1085930}{Frequency-response
  masking approach for the synthesis of sharp linear phase digital filters},
  IEEE transactions on circuits and systems 33~(4) (1986) 357--364.
\newline\urlprefix\url{https://ieeexplore.ieee.org/document/1085930}

\bibitem{lee2006unified}
W.~R. Lee, L.~Caccetta, K.~L. Teo, V.~Rehbock,
  \href{https://ieeexplore.ieee.org/document/1677911}{A unified approach to
  multistage frequency-response masking filter design using the wls technique},
  IEEE transactions on signal processing 54~(9) (2006) 3459--3467.
\newline\urlprefix\url{https://ieeexplore.ieee.org/document/1677911}

\bibitem{vaidyanathan2006multirate}
P.~P. Vaidyanathan, Multirate systems and filter banks, Pearson Education
  India, 2006.

\bibitem{sakthivel2018low}
V.~Sakthivel, E.~Elias,
  \href{https://www.sciencedirect.com/science/article/abs/pii/S0045790617334390}{Low
  complexity reconfigurable channelizers using non-uniform filter banks},
  Computers \& Electrical Engineering 68 (2018) 389--403.
\newline\urlprefix\url{https://www.sciencedirect.com/science/article/abs/pii/S0045790617334390}

\end{thebibliography}
\end{document}